\documentclass[12pt]{article}
\usepackage{epsfig}

\usepackage{graphicx}
\usepackage[centertags]{amsmath}
\usepackage{amsfonts}
\usepackage{amssymb}
\usepackage{amsthm}
\usepackage{newlfont}
\usepackage{epsfig}
\usepackage{amscd}

\newcommand{\G}{\Gamma}

\theoremstyle{plain}

\theoremstyle{definition}

\theoremstyle{remark}
\newcounter{one}
\begin{document}
\title{An integrable generalization  of the Toda law \\ to the square lattice}
\author{
P.M. Santini\thanks{Dipartimento di Fisica, Universit\`a di Roma 
``La Sapienza'' and
Istituto Nazionale di Fisica Nucleare, Sezione di Roma,
Piazz.le Aldo Moro 2, I--00185 Roma, Italy
e-mail: paolo.santini@roma1.infn.it}, 
M. Nieszporski\thanks{
Katedra Metod Matematycznych Fizyki,
Uniwersytet Warszawski
ul. Ho\.za 74, 00-682 Warszawa, Poland
 e-mail: maciejun@fuw.edu.pl,
tel: +48 22 621 77 57,
fax: +48 22 622 45 08},~and
A. Doliwa\thanks{Uniwersytet Warmi\'{n}sko-Mazurski w Olsztynie,
Wydzia{\l} Matematyki i Informatyki,
ul.~\.{Z}o{\l}nierska 14 A, 10-561 Olsztyn, Poland,
e-mail: {\tt doliwa@matman.uwm.edu.pl}}~
}
\maketitle

\begin{abstract}
We generalize the Toda lattice (or Toda chain) 
equation to the square lattice; i.e., we construct an integrable 
nonlinear equation, for a scalar field taking values on the square lattice 
and depending on a continuous (time) variable, characterized by an exponential law of interaction in 
both discrete directions of the square lattice.  We construct the 
Darboux-Backl\"und 
transformations for such lattice, and the corresponding 
formulas describing their superposition. We finally use these 
Darboux-Backl\"und transformations to generate examples of 
explicit solutions of exponential and rational type. The exponential 
solutions describe the evolution of one and two smooth two-dimensional 
shock waves on the square lattice.

\end{abstract}

\section{Introduction}

The Toda lattice \cite{Toda1,Toda2,Toda3}
\begin{equation} 
\label{1d-toda}
\frac{d^2Q_m}{dt^2}=\Delta_me^{\Delta_mQ_{m-1}},
\end{equation}
where $\Delta_mf_m=f_{m+1}-f_m$ is the difference operator  
and $Q=Q_m(t)$ is a 
dynamical function on a one dimensional lattice ($m\in{\mathbb Z}$), 
is one of the most famous integrable nonlinear lattice equations. 
It describes the 
dynamics of a one - dimensional physical lattice whose masses are 
subjected to an interaction potential of 
exponential type. 
The infinite, finite and periodic Toda lattice (\ref{1d-toda}), as well as 
its numerous extensions
\cite{Hir,Levi,Date,Hirota2,Mikhailov,Bruschi,Rui,Sur,Yam}, 
have applications in various other physical and 
mathematical contexts \cite{Hiro,HiroS,Perk,Gib,Lu1,Lu2,Lukas}.

In this paper we introduce the following integrable 
generalization of the Toda law (\ref{1d-toda}) (i.e., the law characterized by an exponential 
interaction between nearest neighbours) to a 2-dimensional lattice:
 \begin{eqnarray}
\label{2d-toda}
\begin{array}{l}
C_{m,n}\frac{d}{dt}\!\left(\!
\frac{1}{C_{m,n}}\!\frac{dQ_{m,n}}{dt}\!
\right)\!=\! 
\Delta_m\!\left(\!C_{m,n}\!C_{m-1,n}e^{\Delta_mQ_{m\!-\!1,\!n}}\right)\!+\! 
\Delta_n\!\left(\!C_{m,n}\!C_{m,\!n\!-\!1}e^{\Delta_n\!Q_{m,n\!-\!1}}\right), \\
~~ \\
\frac{C_{m+1,n+1}}{C_{m,n}}=e^{-\Delta_m\Delta_nQ_{m,n}},
\end{array}
\end{eqnarray}
where 
$Q=Q_{m,n}(t)$ and $C=C_{m,n}(t)$ are  
dynamical functions on the square lattice ($(m,n)\in{\mathbb Z}^2$) and
$
\Delta_mf_{m,n}=f_{m+1,n}-f_{m,n},~~~\Delta_nf_{m,n}=f_{m,n+1}-f_{m,n}
$ are the difference operators in the $m$ and $n$ directions.

Starting with the linear 5-point scheme (\ref{5p-dSchr}) \cite{NSD}, in \S 2 we 
construct a Lax pair  for eq. (\ref{2d-toda}); 
in \S 3 we construct the 
Darboux-B\"acklund Transformations (DBTs) for this 2D Toda lattice, 
and the corresponding formulas describing the superposition of such DBTs; 
in \S 4 we use these transformations to construct explicit solutions 
of exponential and rational type of the 2+1 dimensional Toda lattice (\ref{2d-toda}).

We remark that, in the literature related to integrable systems, there exist already 
three 2+1 dimensional generalizations of the Toda lattice (\ref{1d-toda}). The first  
one  is  obtained 
replacing the second derivative $d^2/dt^2$ in (\ref{1d-toda}) by the hyperbolic operator $\partial^2/\partial x\partial y$ 
\cite{Mikhailov} (see also \cite{Laplace,Darboux}), or by the elliptic operator $\partial^2/\partial z\partial 
\bar z$. 
In this equation, therefore, 
the scalar field $Q$ depends on two continuous variables $x,y$ and on one discrete 
variable $m$: $Q_m(x,y)$. The second generalization \cite{Levi} can be viewed as a variant of the first, 
in which one of the two continuous variables, say $x$, is suitably discretized. The third 
generalization \cite{Date,Hirota2} is obtained discretizing both $x$ and $y$ variables. 
In the generalization (\ref{2d-toda}) we propose in this paper, instead, the scalar field $Q$ 
depends on the continuous time variable $t$, through the Sturm-Liouville operator in the left hand side of (\ref{2d-toda}), 
and on the two discrete indeces $(m,n)\in{\mathbb Z}^2$ of the square lattice ($Q=Q_{m,n}(t)$), through  the 
exponential law of interaction between nearest neighbours in both $m$ and $n$ directions. 

We remark that a 2+1 dimensional generalization of the Volterra system on the square 
lattice has recently appeared  \cite{Hu}.

\section{The 2D generalization of the Toda lattice}
The Lax pair (zero curvature representation) for the Toda lattice   
can be written in the following form \cite{Flaschka},\cite{Manakov}:
\begin{eqnarray}
\label{1d-Lpair}
\begin{array}{l}
\frac{\Gamma_{m}}{\Gamma_{m+1}}\phi_{m+1}+\frac{\Gamma_{m-1}}{\Gamma_{m}}\phi_{m-1}-
F_{m}\phi_{m}=\lambda\phi_{m} , \\
\frac{d\phi_{m}}{dt}=\frac{\Gamma_{m}}{\Gamma_{m+1}}\phi_{m+1}-
\frac{\Gamma_{m-1}}{\Gamma_{m}}\phi_{m-1}, 
\end{array}
\end{eqnarray} 
where $\lambda$ is the constant eigenvalue of the self-adjoint 3-point scheme (\ref{1d-Lpair}a), 
$\Gamma_{m}(t),F_{m}(t)$ are dynamical functions on the lattice, 
the eigenfunction $\phi_{m}(t,\lambda)$ solves simultaneously the Lax pair (\ref{1d-Lpair}), and the Toda field $Q$ 
is related to $\G,F$ in the following way
\begin{equation}
F_{m}=-\frac{dQ_{m}}{dt},~~~~~\Gamma_{m}=e^{-\frac{Q_{m}}{2}}.
\end{equation}

A key progress towards the 
generalization of the  Toda law  (\ref{1d-toda}) to a 2-dimensional lattice
has been recently made in \cite{NSD}; in that paper, devoted to the investigation 
of discretizations of elliptic operators on 2D lattices admitting Darboux 
transformations (DTs), the following results were, in particular, established. 
\begin{itemize}
\item 
The linear and self-adjoint 5-point scheme on the star of the square lattice:
\begin{eqnarray}
\label{5p}
\begin{array}{l}
{\mathcal L}_{5}\tilde\psi_{m,n}:=a_{m,n}\tilde\psi_{m+1,n}+a_{m-1,n}\tilde\psi_{m-1,n}+ \\
b_{m,n}\tilde\psi_{m,n+1}+b_{m,n-1}\tilde\psi_{m,n-1}-f_{m,n}\tilde\psi_{m,n}=0,
\end{array}
\end{eqnarray}
a natural discretization of the self-adjoint elliptic (if $AB>0$) operator 
\begin{equation}
(A\Psi_x)_x+(B\Psi_y)_y={\mathcal D} \Psi
\end{equation}
in canonical form, admits DTs.
\item 
The 5-point scheme (\ref{5p}) admits a distinguished gauge equivalent form:
\begin{eqnarray}
\label{5p-dSchr} 
\begin{array}{l}
{\mathcal L}_{SchInt}\psi_{m,n}:=
\frac{\Gamma_{m,n}}{\Gamma_{m+1,n}}\!\psi_{m+1,n}\!+\!\frac{\Gamma_{m-1,n}}{\Gamma_{m,n}}\!\psi_{m-1,n}\!+ \\
\frac{\Gamma_{m,n}}{\Gamma_{m,n+1}}\!\psi_{m,n+1}\!+\!\frac{\Gamma_{m,n-1}}{\Gamma_{m,n}}\!\psi_{m,n-1}\!-\!
F_{m,n}\!\psi_{m,n}=0,
\end{array}
\end{eqnarray}
obtained from (\ref{5p}) via the following gauge transformation:
\begin{equation}
\label{gaugeSch-1}
{\mathcal L}_{SchInt}=\frac{g_{m,n}}{\Gamma_{m,n}}
{\mathcal L}_{5}\frac{g_{m,n}}{\Gamma_{m,n}},
\end{equation}
with $F_{m,n}=f_{m,n}(g^2_{m,n}/\Gamma^2_{m,n})$, where $g$ and $\Gamma$ are defined by:
\begin{equation}
\label{gaugeSch-2}
a_{m,n}g_{m+1,n}=b_{m,n}g_{m+1,n},~~~\Gamma_{m,n}=
\sqrt{a_{m,n}g_{m,n}g_{m+1,n}}=\sqrt{b_{m,n}g_{m,n}g_{m,n+1}}.
\end{equation}

\end{itemize}

The linear problem (\ref{5p-dSchr}), a natural 2D generalization of the linear problem 
(\ref{1d-Lpair}a), 
satisfies the following basic properties: i) it possesses DTs 
(inherited from the DTs of (\ref{5p})); ii) it reduces, 
in the natural continuous limit, to the 2D Schr\"odinger equation: 
\begin{equation}
\label{2d-schr}
\psi_{xx}+\psi_{yy}+u(x,y)\psi=0.
\end{equation}
For these two reasons the self-adjoint 5-point scheme (\ref{5p-dSchr}) was identified 
in \cite{NSD} as a proper ''integrable'' discretization of the 2D Schr\"odinger equation, a good  
starting point in the search for  
integrable discretisations of the nonlinear symmetries associated 
with the spectral problem (\ref{2d-schr}) 
 and in the search for an integrable 
generalization of the Toda equation to a square lattice. 

The 2 dimensional generalization of the Lax  pair (\ref{1d-Lpair}) proposed in this 
paper is indeed based on the linear problem (\ref{5p-dSchr}), and reads
\begin{eqnarray}
\label{2d-Lpair}
\begin{array}{l}
\frac{\Gamma_{m,n}}{\Gamma_{m+1,n}}\psi_{m+1,n}+\frac{\Gamma_{m-1,n}}{\Gamma_{m,n}}\psi_{m-1,n}+
\frac{\Gamma_{m,n}}{\Gamma_{m,n+1}}\psi_{m,n+1}+\frac{\Gamma_{m,n-1}}{\Gamma_{m,n}}\psi_{m,n-1}=
F_{m,n}\psi_{m,n}, \\
~~ \\
\frac{d\psi_{m,n}}{dt}=\frac{C_{m,n}}{2}\left[\frac{\Gamma_{m,n}}{\Gamma_{m+1,n}}\psi_{m+1,n}-
\frac{\Gamma_{m-1,n}}{\Gamma_{m,n}}\psi_{m-1,n}+
\frac{\Gamma_{m,n}}{\Gamma_{m,n+1}}\psi_{m,n+1}-\frac{\Gamma_{m,n-1}}{\Gamma_{m,n}}\psi_{m,n-1}\right].
\end{array}
\end{eqnarray} 
It is easy to verify that this system of linear equations for the 
eigenfunction $\psi_{m,n}(t)$ 
is compatible iff the coefficients $\Gamma,F,C$ satisfy 
the following nonlinear equations:
\begin{eqnarray}
\label{nlin}
\begin{array}{l}
\frac{dF_{m,n}}{dt}\! =\!C_{m+1,n}\! \left( \frac{\G_{m,n}}{\G_{m+1,n}} \right)^2\!\!\! 
\!-\!C_{m-1,n}\! \left( \frac{\G_{m-1,n}}{\G_{m,n}} \right)^2\!\!\!
+\! C_{m,n+1}\! \left( \frac{\G_{m,n}}{\G_{m,n+1}} \right)^2\!\!\!
-\! C_{m,n-1}\! \left( \frac{\G_{m,n-1}}{\G_{m,n}} \right)^2\!\!\!, \\
~~  \\
\frac{d\G_{m,n}}{dt} = \frac{1}{2}C_{m,n} F_{m,n}\G_{m,n}, \\
~~  \\
\frac{ C_{m+1,n+1} }{ C_{m,n} } = \left( \frac{ \G_{m+1,n+1} \G_{m,n} }{ \G_{m+1,n} \G_{m,n+1} } \right)^2.
\end{array}
\end{eqnarray}
Equations (\ref{nlin})   
can be conveniently rewritten as the 2+1 dimensional generalization (\ref{2d-toda}) 
of the exponential interaction law of Toda,  
in terms of the scalar field $Q_{m,n}(t)$ defined by
\begin{equation}
F_{m,n}C_{m,n}=-\frac{dQ_{m,n}}{dt},~~~~~\Gamma_{m,n}=e^{-\frac{1}{2}Q_{m,n}}.
\end{equation}

In the one dimensional limit in which $\psi_{m,n}(t)$ depends trivially on 
$n$ and the coefficients $\Gamma,F,C$ do not depend on $n$:
\begin{equation}
\psi_{m,n}(t)=\phi_m(t)(-z)^n,~~\Gamma_{m,n}=\Gamma_m,~~F_{m,n}=F_m,~C_{m,n}=C_m,
\end{equation}
it follows from (\ref{2d-toda}b) that $C$ is an arbitrary function of $t$ independent of $m$:  
$C_{m}(t)=C(t)$ and equation (\ref{2d-toda}a) reduces to 
\begin{equation}  
\frac{1}{C(t)}\frac{d}{dt}\left(
\frac{1}{C(t)}\frac{dQ_{m}}{dt}
\right)=\Delta_me^{\Delta_mQ_{m-1}}.
\end{equation} 
Suitably rescaling $t$ (or choosing $C(t)=1$), one finally recovers the Toda lattice (\ref{1d-toda}) and 
its Lax pair (\ref{1d-Lpair}), with $\lambda=z+z^{-1}$.

We first remark that the 5-point schemes (\ref{2d-Lpair}) on the star of the square lattice are 
the simplest and natural generalization of the 3-point schemes (\ref{1d-Lpair}). Therefore 
we expect that our Toda 2D lattice (\ref{2d-toda}) be the simplest integrable generalization 
of (\ref{1d-toda}) on the square lattice. 

We also remark that the Toda 2D-lattice system (\ref{2d-toda}) can be  
rewritten as a single equation, observing that equation (\ref{2d-toda}b) is 
identically satisfied by the following parametrization
\begin{equation}
\label{tau-parametrization}
\Gamma^2_{m,n}=\frac{\tau_{m,n}}{\tau_{m+1,n+1}},~~~C_{m,n}=\frac{\tau_{m+1,n}\tau_{m,n+1}}{\tau_{m,n}\tau_{m+1,n+1}}
\end{equation}
in terms of the single scalar field $\tau_{m,n}$. Then the system  (\ref{2d-toda}) reduces to the single equation 
\begin{eqnarray}
\label{tau-equation}
\begin{array}{l} 
W[\!\tau_{m+1,n}\!\tau_{m,n+1},W[\tau_{m+1,n+1},\!\tau_{m,n}]]\!=\!
\tau^2_{m+1,n}\left(\tau_{m,n}\!\tau_{m,n+2}\!-\!\tau_{m+1,n+1}\!\tau_{m-1,n+1}\right)+ \\ 
\tau^2_{m,n+1}\left(\tau_{m,n}\!\tau_{m+2,n}\!-\!\tau_{m+1,n+1}\!\tau_{m+1,n-1}\right),
\end{array}
\end{eqnarray}
where 
\begin{equation}
W[\alpha,\beta]:=\alpha\dot\beta - \dot\alpha\beta
\end{equation}
is the usual Wronskian of the two functions $\alpha$ and $\beta$. It turns out \cite{DGNS} 
that the scalar function $\tau_{m,n}(t)$ is related to the $\tau$-function of the BKP 
hierarchy \cite{Miwa}; therefore equation (\ref{tau-equation}) gives the $\tau$-function 
formulation of the 2+1 dimensional Toda system (\ref{2d-toda}).

If the $\tau$-function depends only on $m$: $\tau_{m,n}(t)=\tau_m(t)$, equation (\ref{tau-equation}) 
reduces to equation
\begin{equation}
\tau_{m+1}^2 H[\tau_m] - \tau_{m}^2 H[\tau_{m+1}] = 0, 
\end{equation}
where
\begin{equation}
H[\tau_m]:=\ddot\tau_{m} \tau_{m} - {\dot\tau}^2_{m}+\tau_{m+1}\tau_{m-1} - \tau^2_{m},
\end{equation} 
implying that 
\begin{equation}
H[\tau_m]=f(t)\tau^2_{m},  
\end{equation}
with $f(t)$ arbitrary function of $t$. By the change of variable $\tilde\tau_m(t)=\break
exp(-y(t))\tau_m(t)$, with 
$\ddot y(t)=f(t)$, we recover the $\tau$-function formulation 
\cite{Hir} $H[\tilde\tau_m]=0$ of the Toda lattice (\ref{1d-toda}).

We finally remark that $F$ and $C$ constants and $\Gamma=e^{\frac{CF}{2}t}$ is 
the trivial solution of (\ref{nlin}); correspondingly: $Q=-FCt$ blows linearly in time. 
Solutions which are perturbations of this solution will exhibit such a linear blow up 
in time, which can be removed introducing the change of variables $Q_{m,n}=P_{m,n}-FCt+\delta$. 
Then equation (\ref{2d-toda}) becomes:

\begin{eqnarray}
\label{2d-todap}
\begin{array}{l}
C_{m,n}\frac{d}{dt}\!\left(\!
\frac{1}{C_{m,n}}\! \left( \frac{dP_{m,n}}{dt}\!  +\!FC\! \right) \!
\right)\!=\! 
\Delta_m\!\left(\!C_{m,n}\!C_{m-1,n}e^{\Delta_m P_{m-1,n}}\right)
\!+ \\ 
+\Delta_n\!\left(\!C_{m,n}\!C_{m,n-1}e^{\Delta_n P_{m,n-1}}\right),~~~~~~~
\frac{C_{m+1,n+1}}{C_{m,n}}=e^{-\Delta_m\Delta_n P_{m,n}}.
\end{array}
\end{eqnarray}

We end this section remarking that algebro-geometric solutions of the eigenvalue problem for a generic 
5-point scheme were constructed in \cite{Kri2}.

\section{Darboux-Backl\"und transformations and their superposition}

It is a straightforward (but long) calculation to verify that the  DBTs  
for the 2D generalization of the Toda lattice (\ref{nlin}), (\ref{2d-toda}) read as follows:

\begin{eqnarray}
\begin{array}{l}
\label{DB1}
\Delta_m     \left( \frac{\G'}{\G} \theta  \psi ' \right) _{m,n}=
\frac{\G_{m,n-1}}{\G_{m,n}} \theta_{m,n} \theta_{m,n-1} 
\tilde{\Delta}_{-n} \frac{\psi_{m,n}}{ \theta_{m,n}},
\\
\Delta_n     \left( \frac{\G'}{\G} \theta  \psi '  \right)_{m,n}=
-\frac{\G_{m-1,n}}{\G_{m,n}} \theta_{m,n} \theta_{m,n-1}
\tilde{\Delta}_{-m}  \frac{\psi_{m,n}}{ \theta_{m,n}},
\\
\frac{d}{dt} \left( \frac{\G'}{\G} \theta  \psi '  \right) _{m,n}=
 C_{m,n} \frac{\G_{m-1,n}\G_{m,n-1}}{\G_{m,n}^2} \theta_{m-1,n} \theta_{m,n-1} 
 \left( \frac{\psi_{m-1,n}}{ \theta_{m-1,n}}
- \frac{\psi_{m,n-1}}{ \theta_{m,n-1}} \right),
\end{array}
\end{eqnarray}
\begin{eqnarray}
\label{DB2}
\begin{array}{l}
\G'\, ^2_{m,n}=(\G \theta) _{m-1,n-1} \frac{\G_{m,n}}{\theta_{m,n}}, \\
C_{m,n}'=\frac{(\G \theta)_{m-1,n} (\G \theta)_{m,n-1}}{(\G \theta)_{m,n} 
(\G \theta)_{m-1,n-1}} C_{m,n}, \\
F'_{m,n}=(\G \theta)_{m-1,n-1} \left[ \frac{1}{(\G \theta)_{m-1,n}}+\frac{1}{(\G \theta)_{m,n-1}}\right]
+\frac{\theta_{m,n}}{\G_{m,n}} \left[ \left( \frac{\G}{\theta} \right)_{m-1,n}+\left( \frac{\G}{\theta} \right)_{m,n-1}\right],
\end{array}
\end{eqnarray}
where $\tilde{\Delta}_{-m} f_{m,n}=f_{m,n}-f_{m-1,n}$ and $\tilde{\Delta}_{-n}=f_{m,n}-f_{m,n-1}$.  
In these equations: $\theta_{m,n}(t)$ is a solution of the Lax pair (\ref{2d-Lpair}), 
for the coefficients $\Gamma_{m,n},F_{m,n},C_{m,n}$;  
$\psi'_{m,n}(t)$ in (\ref{DB1}) 
is the transformed (via $\theta_{m,n}$) solution of the Lax pair (\ref{2d-Lpair}),  
for the transformed (via (\ref{DB2})) coefficients $\Gamma'_{m,n},F'_{m,n},C'_{m,n}$. 

The so-called spatial part of the above DBTs  
was already written in \cite{NSD}; the temporal part, describing the time dependence of the transformed 
solution $\psi'_{m,n}(t)$ of (\ref{2d-Lpair}), and the transformation 
law (\ref{DB2}b) for the coefficient $C_{m,n}(t)$, are new results of this paper. 

It is well-known \cite{Bianchi} that it is possible to combine DBTs of a given integrable system, to construct 
superposition formulas and a permutability 
diagram of DBTs. For the 2D Toda lattice (\ref{nlin}),(\ref{2d-toda}) it is possible to prove the following result 
(see also \cite{DGNS} for the Bianchi permutability diagram of the general self-adjoint scheme on the star of 
the square lattice).

\vskip 5pt
\noindent
Consider a solution ($\Gamma_{m,n},F_{m,n},C_{m,n}$) of the 2D Toda lattice (\ref{nlin}),(\ref{2d-toda}), and 
let $\theta^{(1)}_{m,n}$ and $\theta^{(2)}_{m,n}$ 
be two independent solutions of the Lax pair (\ref{2d-Lpair}), corresponding to the coefficients 
$\Gamma_{m,n},F_{m,n},C_{m,n}$. Superimposig the two DBTs (\ref{DB1}) with respect to $\theta^{(1)}_{m,n}$ 
and $\theta^{(2)}_{m,n}$, 
one obtains the new solution ($\Gamma^{(12)}_{m,n},F^{(12)}_{m,n},C^{(12)}_{m,n}$) of the nonlinear 
system (\ref{nlin})-(\ref{2d-toda}) through the following formulas: 
\begin{equation}
\label{permutability}
\begin{array}{l}
(\Gamma^{(12)}_{m+1,n+1})^2=(\Gamma_{m,n})^2 \frac{\Sigma_{m,n}}{\Sigma_{m+1,n+1}},
\\
\\
F^{(12)}_{m+1,n+1}=F_{m,n}+
\frac{1}{\Sigma_{m,n+1}}  \frac{\Gamma_{m-1,n}}{\Gamma_{m,n+1}}
(\theta ^{(1)}_{m-1,n} \theta ^{(2)}_{m,n+1}-\theta ^{(1)}_{m,n+1} \theta ^{(2)}_{m-1,n})
+\\ \\
\frac{1}{\Sigma_{m+1,n}} \frac{\Gamma_{m,n-1}}{\Gamma_{m+1,n}}
(\theta ^{(1)}_{m+1,n} \theta ^{(2)}_{m,n-1}-\theta ^{(1)}_{m,n-1} \theta ^{(2)}_{m+1,n})=
\\
\\
=\frac{\Sigma_{m,n}\Sigma_{m+1,n+1}}{\Sigma_{m+1,n}\Sigma_{m,n+1}}
[ F_{m,n}+ \frac{1}{\Sigma_{m+1,n+1}} \frac{\Gamma^2_{m,n}}{\Gamma_{m+1,n}\Gamma_{m,n+1}}
(\theta ^{(1)}_{m,n+1} \theta ^{(2)}_{m+1,n}-\theta ^{(1)}_{m+1,n} \theta ^{(2)}_{m,n+1})+
\\
\\
\frac{1}{\Sigma_{m,n}} \frac{\Gamma_{m-1,n}\Gamma_{m,n-1}}{\Gamma^2_{m,n}}
(\theta ^{(1)}_{m,n-1} \theta ^{(2)}_{m-1,n}-\theta ^{(1)}_{m-1,n} \theta ^{(2)}_{m,n-1})
 ]
\\
\\
C^{(12)}_{m+1,n+1}= \left(
\frac{\Gamma_{m+1,n} \Gamma_{m,n+1}}{\Gamma_{m,n}
\Gamma_{m+1,n+1}} \right)^2
\frac{\Sigma_{m+1,n} \Sigma_{m,n+1}}{\Sigma_{m,n}
\Sigma_{m+1,n+1}} C_{m+1,n+1},
\end{array}
\end{equation}
where the function $\Sigma_{m,n}$ is obtained  integrating  the first order compatible equations 
\begin{equation}
\label{Sigma}
\begin{array}{l}
\frac{d}{dt} \Sigma_{m,n}=  C_{m,n} 
\frac{\Gamma_{m-1,n}\Gamma_{m,n-1}}{\Gamma^2_{m,n}}
(\theta ^{(1)}_{m,n-1}\theta ^{(2)}_{m-1,n}
-\theta ^{(1)}_{m-1,n}\theta ^{(2)}_{m,n-1})\\
\Sigma_{m+1,n}-\Sigma_{m,n}=\frac{\Gamma_{m,n-1}}{\Gamma_{m,n}}
 (\theta ^{(1)}_{m,n-1} \theta ^{(2)}_{m,n}-\theta ^{(1)}_{m,n} \theta ^{(2)}_{m,n-1}),\\
\Sigma_{m,n+1}-\Sigma_{m,n}=\frac{\Gamma_{m-1,n}}{\Gamma_{m,n}}
 (\theta ^{(1)}_{m,n} \theta ^{(2)}_{m-1,n}-\theta ^{(1)}_{m-1,n} \theta ^{(2)}_{m,n}).
\end{array}
\end{equation}
This scheme is often used to construct the two-soliton solution 
knowing the one-soliton solution of the system (in this case $\theta^{(1)}_{m,n}$ 
and $\theta^{(2)}_{m,n}$ 
are the eigenfunctions of the one-soliton solution corresponding to two different sets of parameters). 

\section{Solutions of exponential and rational type}

The existence of DBTs is considered one of 
the basic properties of an integrable nonlinear 
system. In particular, it allows one to construct iteratively solutions from 
simpler solutions, via an endless procedure. 

In this section we show some example of explicit solutions 
of exponential and rational type of the 2D Toda lattice (\ref{2d-toda}), obtained using the 
DBTs (\ref{DB2}).

We consider as starting solution of the system (\ref{nlin}) the trivial one, corresponding to $F,C$ constants 
and $\Gamma=e^{\frac{FC}{2}t}$ (for this solution $Q=-FCt$) and, correspondingly, we look for an exponential  
solution of the Lax pair (\ref{2d-Lpair}): 
\begin{equation}
\label{exp}
\psi_{m,n}(t)=e^{\alpha m+\beta n+\omega t+\delta},
\end{equation} 
obtaing for the coefficients $\alpha,\beta,\omega$ the following equations:
\begin{eqnarray}
\label{constr}
\begin{array}{c}
\cosh\alpha+\cosh\beta=\frac{F}{2}, \\
\sinh\alpha+\sinh\beta=\frac{\omega}{C}.
\end{array}
\end{eqnarray} 
These equations can be interpreted in the following way: given the constants $F,C$, 
equation (\ref{constr}a) establishes a constraint between the ``wave numbers'' $\alpha$ and $\beta$; 
once this constraint is satisfied, equation (\ref{constr}b) gives the ``dispersion relation'' 
$\omega$ in terms of $\alpha$ and $\beta$. 

Looking for real and non singular solutions of exponential type, in the following we restrict our analysis to the 
case $F,C,\alpha,\beta,\delta\in{\mathbb R}$, postponing the study of other possible choices to a subsequent work. Then  
equation (\ref{constr}) implies that $F\ge 4$ and that 
the parameters $\alpha$ and $\beta=\cosh^{-1}(\frac{F}{2}-\cosh\alpha )$ must range in the interval  
\begin{equation}
1<\cosh\alpha,\cosh\beta \le\frac{F}{2}-1,~~~F\ge 4
\end{equation}    
(if $F=4$, then $\alpha=\beta=0$). 

We consider now the following solution of (\ref{2d-Lpair}): 
\begin{eqnarray}
\begin{array}{l}
\theta_{m,n}(t)=\cosh\Theta^+_{m,n}(t)+\rho\cosh\Theta^-_{m,n}(t), \\
\Theta^{\pm}_{m,n}(t):=\alpha m\pm\beta n+\omega^{\pm} t+\delta^{\pm},~~
\omega^{\pm}:=C(\sinh\alpha\pm\sinh\beta),
\end{array}
\end{eqnarray} 
consisting of a suitable combination of four exponentials of the type (\ref{exp}), where $\alpha,\beta$ 
satisfy the constraint (\ref{constr}a), and $\rho\ge 0,~\delta^{\pm}\in{\mathbb R}$.

Applying the DBTs (\ref{DB2}) to this basic solution, one obtains the 
following dressed solution of the 2D Toda lattice (\ref{nlin}),(\ref{2d-toda}):
\begin{eqnarray}
\label{2-shock}
\begin{array}{l}
{{\Gamma'}^2_{m,n}}=e^{-Q'_{m,n}(t)}=
\frac{\cosh\Theta^+_{m-1,n-1}(t)+\rho\cosh\Theta^-_{m-1,n-1}(t)}{\cosh\Theta^+_{m,n}(t)+\rho\cosh\Theta^-_{m,n}(t)}e^{FCt}, \\
C'_{m,n}(t)=\frac
{[\cosh\Theta^+_{m-1,n}(t)+\rho\cosh\Theta^-_{m-1,n}(t)][\cosh\Theta^+_{m,n-1}(t)+\rho\cosh\Theta^-_{m,n-1}(t)]}
{[\cosh\Theta^+_{m,n}(t)+\rho\cosh\Theta^-_{m,n}(t)][\cosh\Theta^+_{m-1,n-1}(t)+\rho\cosh\Theta^-_{m-1,n-1}(t)]}C.
\end{array}
\end{eqnarray}   
This solution exhibits, depending on the value of $\rho\ge 0$, the following different features.

\vskip 5pt
\noindent
i) If $\rho=0$, we have the simplified expression:
\begin{eqnarray}
\label{1-shock}
\begin{array}{l}
{{\Gamma'}^2_{m,n}}=e^{-Q'_{m,n}(t)}=
\frac{\cosh\Theta^+_{m-1,n-1}(t)}{\cosh\Theta^+_{m,n}(t)}e^{FCt}, \\
C'_{m,n}(t)=\frac
{[\cosh\Theta^+_{m-1,n}(t)][\cosh\Theta^+_{m,n-1}(t)]}
{[\cosh\Theta^+_{m,n}(t)][\cosh\Theta^+_{m-1,n-1}(t)]}C.
\end{array}
\end{eqnarray}
This solution describes a smooth $2D$-shock wave (a kink); the shock front is the    
phase (straight) line $\Theta^+=const$, forming with the $m$ - axis 
the angle $-\theta$, with $\theta=tan^{-1}\frac{\alpha}{\beta}$. This shock wave propagates with speed 
{\bf v}$^+=-\frac{\omega^+}{2}\sin (2\theta)(1/\beta,1/\alpha)$. 
The values of $Q'_{m,n}+FCt$ ahead and behind the shock front are respectively 
$-(\alpha+\beta)$ and $\alpha+\beta$; then the 
shock strength is $2(\alpha+\beta)$ (see Fig. 1). 

\vskip 5pt
\noindent
ii) If $\rho$ is a finite positive number, the solution describes two smooth $2D$-shock waves 
with the following features. 
The phase (straight) lines $\Theta^{\pm}=const$ form with the $m$ - axis the 
angles $\mp\theta$; they travel with speeds  
{\bf v}$^{\pm}=-\frac{\omega^{\pm}}{2}\sin (2\theta)(1/\beta,\pm 1/\alpha)$ and, consequently, 
their intersection point $P$ travels with constant 
speed {\bf v}$_P=-C(\frac{\sinh\alpha}{\alpha},\frac{\sinh\beta}{\beta})$. In this situation the two 
shock fronts do not coincide, as before, with the phase lines; they are now parallel to the $m$ and $n$ axes and 
intersect in P, dividing the $(m,n)$ plane in the usual 4 quadrants. The values of $Q_{m,n}+FCt$ in the first, 
second, third and fourth quadrants are, respectively, $\alpha+\beta$, $-\alpha+\beta$, $-(\alpha+\beta)$, and 
$\alpha-\beta$ (see Fig. 2). 

\vskip 5pt
\noindent
iii) If $\rho$ is a very small positive parameter: $0<\rho<<1$, the previous two regimes combine in the following 
way. In the finite 
$(m,n)$ plane (or, more precisely, in an inner region of the order $O(\ln(1/\rho)))$, the term $\rho\cosh\Theta^-$ is 
negligeable and the expression (\ref{1-shock}) is a good approximation of the solution, which then describes the 
single transversal shock wave of the regime i). In the outer region, that term is not negligeable anymore 
and the regime ii) becomes dominant. Both ends of the transversal shock front bifurcate into two semilines 
parallel to the $m$ and $n$ axes (see Fig.s 3 and 4). 
One could actually say that the regime iii) is the generic one; but, for $\rho=O(1)$, 
the inner region is not visible, since it is smaller than a single elementary square of the square 
lattice $(m,n)$. The inner region is visible if it 
contains at least one elementary square of the lattice. If the spacing of the square lattice is $1$, a rough 
extimate for this condition is that $0<\rho<min(1/(|\alpha |e),1/(|\beta |e))$, where $e$ is the Neper constant.  

The possible existence of web-like structures in the inner region, typical of 
2+1 dimensional soliton models \cite{Gino}, will be explored in a subsequent work.

Starting with the trivial solution 
$F=4,~C$ constant ($\Rightarrow~\Gamma=e^{4Ct}$) of the system (\ref{nlin}), it is also possible to construct rational 
solutions. It is straightforward to verify, for instance, that 
\begin{equation}
\theta_{m,n}(t):= mn+(Ct+a)m+(Ct+b)n+C^2t^2+C(a+b)t+d
\end{equation}
is a polynomial solution of the system (\ref{2d-Lpair}), where $a,b,d$ are arbitrary constant 
coefficients. Substituting it into the DBTs (\ref{DB2}), one obtains a (singular) rational solution  
of the 2D-Toda lattice. We remark that    
this solution could have been derived directly from the solution (\ref{2-shock}) for a 
suitable choice of its free parameters.

\vskip 20pt
\noindent
{\bf Acknowledgments}
\vskip 10pt
\noindent
This work was supported by the cultural 
and scientific agreements between 
the University of Roma ``La Sapienza'' 
and the Universities of Warsaw and Olsztyn. 
It was partially supported by KBN grant 2 P03B 126 22.

\vfill
\eject

\begin{figure}
\label{fig:shock1}
\begin{center}
\includegraphics[scale=0.7]{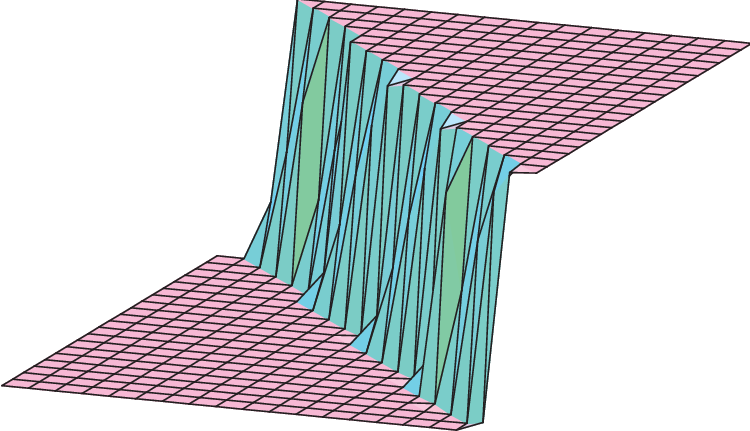}
\caption{If $\rho=0$ and $\alpha,\beta\in{\mathbb R}~(F>4)$, 
the solution $Q_{m,n} +FCt$ in (\ref{2-shock}) describes 
a smooth 2D-shock wave propagating with velocity {\bf v}$^+=-\frac{\omega^+}{2}\sin (2\theta)(1/\beta,1/\alpha)$. 
The shock front is a straight line forming the angle $-\theta$,  
$\theta=tan^{-1}\frac{\alpha}{\beta}$ with the $m$- axis. 
In this figure: 
$\alpha=5,~\beta=4~(F=101),~\delta^{\pm}=1,~C=1,~\rho=0$}
\end{center}
\end{figure}

\begin{figure}
\label{fig:shock2}
\begin{center}
\includegraphics[scale=0.8]{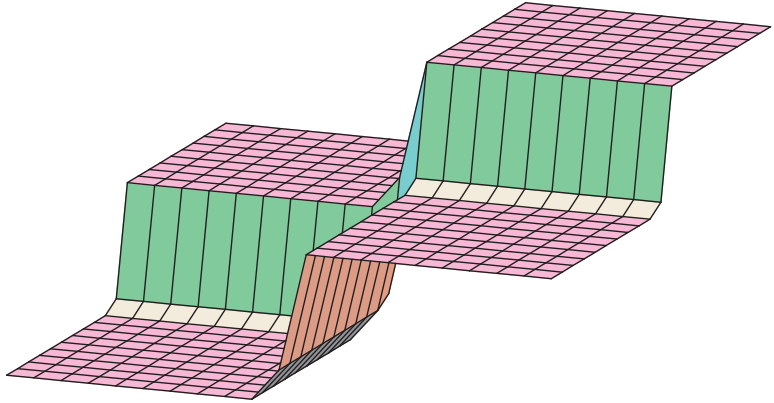}
\caption{For $\alpha,\beta\in{\mathbb R}~(F>4)$ and $\rho=O(1)$, 
the solution $Q_{m,n} +FCt$ in (\ref{2-shock}) describes two 
shock waves with fronts parallel to the $m$ and $n$ axes. The intersection point $P$ of these two 
fronts travels with velocity {\bf v}$_P=-C(\frac{\sinh\alpha}{\alpha},\frac{\sinh\beta}{\beta})$. In this figure: 
$\alpha=5,~\beta=4~(F=101),~\delta^{\pm}=1,~C=1,~\rho=1$}
\end{center}
\end{figure}

\begin{figure}
\label{fig:top3b}
\begin{center}
\includegraphics[scale=0.5]{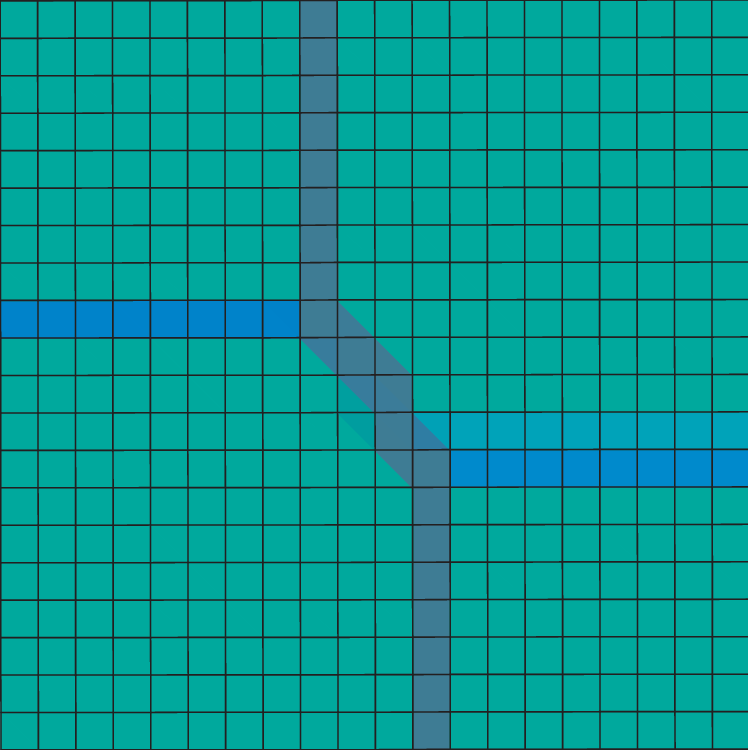}
\caption{A view from the top of the solution for $\rho=10^{-7}$. In the central (finite) region, 
the single shock prevails; this single shock matches with the two orthogonal shocks, which prevail instead in the 
outer region. In this figure: 
$\alpha=5,~\beta=4~(F=101),~\delta^{\pm}=1,~C=1,~\rho=10^{-7}$}
\end{center}
\end{figure}
\begin{figure}
\label{fig:shock3}
\begin{center}
\includegraphics[scale=0.5]{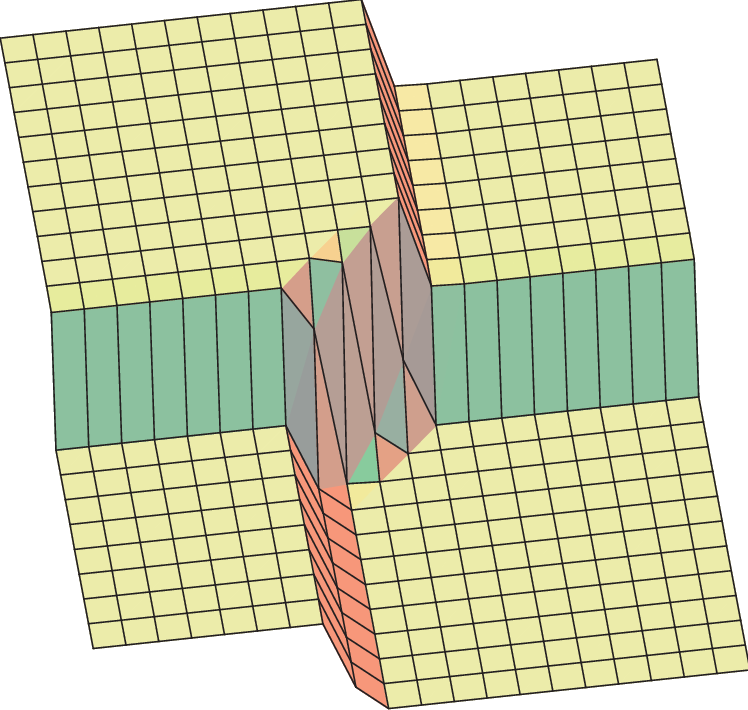}
\caption{A generic view of the solution for $\rho=10^{-7}$. In the central (finite) region, 
the single shock prevails; this single shock matches with the two orthogonal shocks, which prevail  instead in the 
outer region. In this figure: 
$\alpha=5,~\beta=4~(F=101),~\delta^{\pm}=1,~C=1,~\rho=10^{-7}$}
\end{center}
\end{figure}

\end{document}